\documentclass[10pt,a4paper,twocolumn]{IEEEtran}

\usepackage{amssymb}
\usepackage{graphicx}
\usepackage{multirow}
\usepackage{float}
\usepackage{amsmath,amssymb,amsfonts,mathrsfs}
\usepackage{color}
\usepackage{subfig}

\newtheorem{theorem}{\textbf{Theorem}}
\newtheorem{lemma}{\textbf{Lemma}}
\newtheorem{corollary}{\textbf{Corollary}}

\newtheorem{definition}{\textbf{Definition}}

\begin{document}
\renewcommand{\thepage}{11--\arabic{page}}
\setcounter{page}{1}
\setlength{\topsep}{0.0cm}
\setlength{\itemsep}{0.0em}

\title{\Large\bf Constructibility of a Causal/Impulse Free NDS Using Descriptor Form Subsystems} 

\author{Tong Zhou 
\thanks{This work was supported in part by the NNSFC under Grant 61733008, 52061635102 and 61573209.}
\thanks{Tong Zhou is with the Department of Automation and BNRist, Tsinghua University, Beijing, 100084, P.~R.~China
        {(email: {\tt\small tzhou@mail.tsinghua.edu.cn}).}}%
}
\maketitle

\begin{abstract}                          
Requirements are investigated in this paper for each descriptor form subsystem, with which a causal/impulse free networked dynamic system (NDS) can be constructed. For this purpose, a matrix rank based necessary and sufficient condition is at first derived for the causality/impulse freeness of an NDS, in which the associated matrix depends affinely on subsystem connections. From this result, a necessary and sufficient condition is derived for each subsystem, such that there exists a subsystem connection matrix that leads to a causal/impulse free NDS. This condition further leads to a necessary and sufficient condition for the existence of a local static output feedback that guarantees the construction of a causal/impulse free NDS. A prominent property of these conditions are that all the involved numerical computations are performed independently on each individual subsystem, which is quite attractive in reducing computation costs and improving numerical stability for large scale NDS analysis and synthesis. Situations have also been clarified in which NDS causality/impulse freeness is independent of subsystem connections. It has also been made clear that under some situations, local static output feedbacks are not helpful in constructing a causal NDS.
\end{abstract}


\section{Introduction}
\setlength{\itemsep}{0.0em}

In describing plant dynamics, descriptor systems, which are sometimes also called singular systems, have been extensively recognized as an appropriate model. Compared with the well adopted state space model, a descriptor system is believed to be more efficient in keeping structural information of plant dynamics, describing system evolutions with its natural variables, etc. These properties are quite important in analyzing influences of a system parameter on its performances, as well as in understanding responses of its natural variables to external stimulus, etc.  \cite{Dai1989a, Duan2010, yl1985, zlr2015, zz2020}. Similar to the state space model, this model has also been attracting extensive research attentions for a long time, and has been frequently utilized in various fields. Some examples include economy, engineering, biology, etc.

When a descriptor model is adopted in system analysis and synthesis, various particular issues arise. Among them, an essential one is its causality when it is in a discrete time form, which requires that system states and outputs do not depend on its future inputs. A closely related issue in a continuous time descriptor model is that there do not exist impulse terms in the response of its states and outputs to external stimulus \cite{Dai1989a,Luenberger1977,yl1985}. Obviously, requirements on system future inputs are in general not reasonable.  Similarly, existence of impulse modes in a system is usually not greatly appreciated also, as it may significantly deteriorate system performances, and even destroy a system. To avoid occurrence of these undesirable phenomena, significant efforts have been devoted from many researchers and various results have been obtained. For example, some matrix rank based necessary and sufficient conditions are derived in \cite{Dai1989a, it2001} for the causality of a discrete time descriptor system, as well as for the non-existence of impulse modes in a continuous time descriptor system, using the concept of restricted system equivalence. On the basis of graph theory, some necessary conditions, as well as some sufficient conditions, are obtained in \cite{yl1985} for verifying generic causality of a discrete time descriptor system. With the help of the Kronecker canonical form of a matrix pencil, some necessary and sufficient conditions are given in \cite{hm1999} for the causal observability of a continuous time descriptor system, which are also based on matrix ranks. In \cite{zlr2015}, a necessary and a sufficient condition have been derived respectively for the existence of a derivative feedback for the first subsystem guaranteeing impulse controllability/observability of the second subsystem in an interconnected descriptor system consisting of two subsystems. A linear matrix inequality based sufficient condition is given in \cite{wl2018} for the causality of an interconnected system constituted from several uncertain singular subsystems. And so on.

While one of the motivations for introducing a descriptor model is to investigate large scale interconnected systems \cite{Dai1989a, Duan2010,Luenberger1977,Siljak1978,yl1985}, all the above conditions are based on a lumped model, which may not be very suitable when the dimension of the system state vector is large, and/or the system is composed from a large number of subsystems, noting that rank verification for a high dimensional matrix is usually computationally very intensive and numerically quite sensitive \cite{gv1996,Gantmacher1959,hj1991}. On the other hand, limited to our knowledge, there are still no researches on the influences of subsystem connections and parameters on the causality/impulse freeness of a large scale interconnected descriptor system, as well as on the requirements about a subsystem from which a causal/impulse free system can be constructed.

To overcome these difficulties, a matrix rank based necessary and sufficient condition is established in this paper for the causality of a networked dynamic system (NDS) composed of several discrete time descriptor subsystems, as well as for the non-existence of impulse modes in an NDS constituted from several continuous time descriptor subsystems. In this condition,  the associated matrix depends affinely on subsystem connections. A prominent property of this condition is that all the involved numerical computations are performed independently on each individual subsystem. This makes the condition efficient in reducing computation costs and improving numerical stability, scalable for large scale NDS analysis and synthesis, as well as helpful in subsystem parameter selections and NDS topology designs. From this condition, situations have also become clear in which the causality/impulse mode freeness of an NDS is completely and independently determined by its subsystem dynamics, which means that no matter how its subsystem connections are changed, the NDS remains causal/free from impulse modes. In addition, a necessary and sufficient condition on each subsystem has been derived from the aforementioned conclusions, which guarantees the existence of subsystem connections that lead to a causal/impulse free NDS. Furthermore, situations have also been clarified in which a static local output feedback exists that makes the associated subsystem satisfy this condition. It has also been made clear that there exist some situations, under which static local output feedbacks are not helpful in converting a non-causal NDS to causal.

The outline of the remaining of this paper is as follows. At first, in Section II, a descriptor form like model is given for subsystems of an NDS, together with some preliminary results. NDS causality is studied in Section III, together with its non-existence of impulse modes. Section IV investigates requirements on each subsystem such that a causal/impulse free NDS can be constructed through subsystem interactions and local static output feedbacks. Some concluding remarks are given in Section V in which several further issues are discussed. Finally, an appendix is included to give proofs of some technical results.

The following notation and symbols are adopted. ${\cal R}^{m\times n}$ and ${\cal R}^{n}$ stand respectively for the set of $m\times n$ dimensional real matrices and the $n$ dimensional real Euclidean space. ${\rm\bf rank} \left(\cdot\right)$ represents the rank of a matrix, ${\rm\bf null} \left(\cdot\right)$ the (right) null space of a matrix, ${\rm\bf span} \left(\cdot\right)$ the space spanned by the columns of a matrix, while $\cdot^{\perp}$ the matrix whose columns form a base of the (right) null space of a matrix. ${\rm\bf diag}\{X_{i}|_{i=1}^{L}\}$ denotes a block diagonal matrix with its $i$-th diagonal block being $X_{i}$, while ${\rm\bf col}\{X_{i}|_{i=1}^{L}\}$ the vector/matrix stacked by $X_{i}|_{i=1}^{L}$ with its $i$-th row block vector/matrix being $X_{i}$. $I_{n}$, $0_{m}$ and $0_{m\times n}$ represent respectively the $m$ dimensional identity matrix, the $m$ dimensional zero column vector and the $m\times n$ dimensional zero matrix. The subscript is usually omitted if it does not lead to confusions. The superscript $T$ is used to denote the transpose of a matrix/vector.

\section{System Description and Some Preliminaries}

In various actual engineering/biological/economic problems, a plant is usually constituted from several subsystems that may have distinctive dynamics. A plant with this characteristic is usually called an NDS, and a promising method to describe its dynamics is to represent the dynamics of its subsystems with an ordinary model, divide their inputs/outputs into two different classes, which are respectively called external and internal inputs/outputs. With these classifications, subsystem interactions are expressed through transmitting an internal output of one subsystem to some other subsystems as one of their internal inputs.

When subsystem dynamics are linear time invariant (LTI) and subsystem connections are time invariant, this approach has been adopted in \cite{Zhou2015,zyl2018,Zhou2020,zz2020} for investigating NDS regularity/controllability/observability, in which the dynamics of each subsystem is described respectively by a state space model and a descriptor model. With this NDS model, some computationally attractive criteria have been developed respectively for its regularity, controllability and observability. In this paper, this model is adopted once again for studying NDS causality and nonexistence of impulse modes.

More specifically, for an NDS $\bf \Sigma$ constituted from $N$ subsystems, the following model is utilized in this paper to describe the dynamics of its $i$-th subsystem ${\bf{\Sigma}}_i$, in which $i$ belongs to the set $\{1,\;2,\;\cdots,\; N\}$.
\begin{eqnarray}
& &\!\!\!\! \left[\! \begin{array}{c}
{E(i)}{\delta({{x}}(t,i))}\\
{z(t,i)}\\
{{y}(t,i)}
\end{array} \!\right] \nonumber\\
&=& \!\!\!\! \left[\! \begin{array}{ccc}
{A_{\rm\bf xx}{(i)}} & {A_{\rm\bf xv}{(i)}} & {B_{\rm\bf x}{(i)}}\\
{A_{\rm\bf zx}{(i)}} & {A_{\rm\bf zv}{(i)}} & {B_{\rm\bf z}{(i)}}\\
{C_{\rm\bf x}{(i)}} & {C_{\rm\bf v}{(i)}} & {D_{\rm\bf u}{(i)}}
\end{array} \!\right]\left[\! \begin{array}{c}
{x(t,i)}\\
{v(t,i)}\\
{u(t,i)}
\end{array} \!\right]
\label{eqn:1}
\end{eqnarray}
Here, $\delta(\cdot)$ represents either a forward time shift operation or the derivative of a function with respect to time. This implies that the above model can be either continuous time or discrete time. Moreover, $t$ denotes the temporal variable, $x(t,i)$ its state vector, $u(t,i)$ and $y(t,i)$ respectively its external input and output vectors,  while $v(t,i)$ and $z(t,i)$ respectively its internal input and output vectors. An external input/output is also an NDS input/output, while an internal input represents a signal received from another subsystem and an internal output represents a signal sent to some other subsystems.

In order to express interactions among subsystems of the NDS $\rm\Sigma$, denote vectors ${\rm\bf col}\left\{v(t,i)|_{i=1}^{N}\right\}$ and ${\rm\bf col}\left\{z(t,i)|_{i=1}^{N}\right\}$ respectively by $v(t)$ and $z(t)$. Then subsystem interactions can be described by
\begin{equation}
v(t)=\Phi z(t)
\label{eqn:2}
\end{equation}
in which the matrix $\Phi$ is called a subsystem connection matrix (SCM) and assumed to be time invariant.

Throughout this paper, the dimensions of the vectors $x(t,i)$, $z(t,i)$, $v(t,i)$, $u(t,i)$ and $y(t,i)$ of the $i$-th subsystem ${\rm\bf\Sigma}_{i}$, $i=1,2,\cdots,N$, are denoted respectively by $n_{\rm\bf \star}(i)$ with $\star=x$, $z$, $v$, $u$ and $y$. In addition, it is assumed that the matrix $E(i)$ has $n_{\rm\bf e}(i)$ rows. From these dimensions, the dimension becomes clear for each system matrix in the $i$-th subsystem ${\rm\bf\Sigma}_{i}$. It is not required in this paper that $n_{\rm\bf e}(i)=n_{\rm\bf x}(i)$ for each $i=1,2,\cdots,N$. That is, the descriptor form subsystem of Equation (\ref{eqn:1}) is permitted to be rectangular.

It is worthwhile to mention that in addition to represent subsystem connections, the SCM $\Phi$ of Equation (\ref{eqn:2}) is also able to include parameters of a subsystem in the NDS $\rm\bf\Sigma$, provided that the system matrices of that subsystem depend on these parameters through a (generalized) linear fractional transformation. Details can be found in \cite{Zhou2020,zz2020}.

In system analysis and synthesis, a frequently adopted model is the following descriptor system,
\begin{equation}
E{\delta({{x}}(t))}=Ax(t)+Bu(t), \hspace{0.5cm} y(t)=Cx(t)+Du(t)
\label{eqn:3}
\end{equation}
in which $A\in {\cal R}^{m\times n}$, $B\in {\cal R}^{m\times p}$, $C\in {\cal R}^{q\times n}$, $D\in {\cal R}^{q\times p}$ and $E\in {\cal R}^{m\times n}$ are some constant real matrices. When the matrix $E$ is not invertible, this model is sometimes also referred to as a singular system. Compared with the so-called state space model, this model is widely believed to be more natural and more convenient in expressing system constraints and keeping system structure information \cite{Dai1989a,Duan2010,Luenberger1977}.

There are some specific concepts related to a descriptor system. The following definition briefly summarizes those required in this investigation.

\begin{definition}
Assume that a descriptor system is described by Equation (\ref{eqn:3}).
\begin{itemize}
\setlength{\itemsep}{-0.08cm}
\item Its initial state $x(0)$ and input $u(t)$ are called admissible, if there exists at least one trajectory $x(t)$ satisfying this equation.
\item This descriptor system is said to be regular, if $m=n$ and the determinant of the matrix pencil $\lambda E - A$ is not constantly equal to zero.
\item If for each admissible initial state $x(0)$, every solution to Equation (\ref{eqn:3}) with $u(t)\equiv 0$ does not include any impulse,  then this descriptor system is called impulse free, provided that $\delta(\cdot)$ represents the derivative of a function with respect to time.
\item Assume that $\delta(\cdot)$ stands for the forward time shift operation. If at each time instant $t$, the state vector $x(t)$ of Equation (\ref{eqn:3}) is completely determined by its initial state $x(0)$ and its former inputs $u(t)|_{t=0}^{k}$, then this descriptor system is called causal.
\end{itemize}
\label{def:1}
\end{definition}

Regularity is an important concept for descriptor systems. When a descriptor system is not regular, its outputs can not be uniquely determined by its inputs and initial states, even if they are admissible.

Note that when the matrix $E$ is of full column rank (FCR),
the descriptor system of Equation (\ref{eqn:3}) can be easily transformed into a state space model \cite{Dai1989a,hm1999}, which makes its causality/impulse freeness analysis very trivial, and is therefore not attractive in this study. Hence, it is assumed in the remaining of this paper that the matrix $E$ is not of FCR. Under such a situation, the singular value decomposition (SVD) of the matrix $E$ can always be written in the following form,
\begin{displaymath}
E=U_{E}\left[\begin{array}{cc}{\rm\bf diag}\{\sigma_{Ei}|_{i=1}^{r}\} & 0_{r\times (n-r)} \\ 0_{(n-r)\times r} & 0_{(n-r)\times (n-r)} \end{array}\right]V_{E}^{T}
\end{displaymath}
in which $\sigma_{E1} \geq \sigma_{E2} \geq \cdots \geq \sigma_{Er} >0$ and $r$ is equal to the rank of the matrix $E$.  Moreover, $U_{E}$ and $V_{E}$ are respectively $m\times m$ and $n\times n$ dimensional orthogonal matrices. It is now well known that SVD can be performed for every matrix and is numerically quite stable, and the numbers $\sigma_{E1}$, $\sigma_{E2}$, $\cdots$, and $\sigma_{Er}$ are called the singular values \cite{Gantmacher1959,hj1991,gv1996}.

Partition the matrices $U_{E}$ and $V_{E}$ respectively as follows,
\begin{displaymath}
U_{E}=\left[U_{E1}\; U_{E2}\right], \hspace{0.5cm}
V_{E}=\left[V_{E1}\; V_{E2}\right]
\end{displaymath}
in which $U_{E1}\in {\cal R}^{m\times r}$, $U_{E2}\in {\cal R}^{m\times (n-r)}$, $V_{E1}\in {\cal R}^{n\times r}$ and $V_{E2}\in {\cal R}^{n\times (n-r)}$. Using these symbols, we have the following results, which are well known about a descriptor system \cite{Dai1989a,hm1999,it2001}.

\begin{lemma} Let a descriptor system be described by Equation (\ref{eqn:3}).
\begin{itemize}
\setlength{\itemsep}{-0.08cm}
\item Assume that $m=n$ and the descriptor system of Equation (\ref{eqn:3}) is regular, and $\delta(\cdot)$ represents the forward time shift operation. If ${\rm\bf rank} \left([E\;\; B]\right)=m$, then the associated descriptor system is causal if and only if the matrix $U_{E2}^{T}AV_{E2}$ is of FCR.
\item Let $\delta(\cdot)$ represent the derivative of a function with respect to time. Then the associated descriptor system is impulse free, if and only if $m\geq n$ and the matrix $U_{E2}^{T}AV_{E2}$ is of FCR.
\end{itemize}
\label{lemma:1}
\end{lemma}

This lemma reveals that the conditions for the nonexistence of impulse modes in a continuous time descriptor system are actually part of the conditions for the causality of a discrete time descriptor system. This makes it possible to investigate this two problems together.

As regularity verifications have been investigated in \cite{Zhou2020} for the NDS of Equations (\ref{eqn:1}) and (\ref{eqn:2}), it is no longer discussed in this paper. On the other hand, the condition $m\geq n$ can be simply verified. These mean that in NDS causality/impulse freeness verifications, the remaining essential tasks are about the aforementioned two rank conditions. In order to have a concise presentation, these two conditions are called respectively Conditions I and II in the remaining of this paper. More specifically,
\begin{itemize}
\item Condition I: the matrix $U_{E2}^{T}AV_{E2}$ is of FCR.
\item Condition II: ${\rm\bf rank} \left([E\;\; B]\right)=m$.
\end{itemize}

The following results can be obtained straightforwardly from Lemma A.1 of \cite{ac1981}.

\begin{lemma}
Let $A$ and $B$ be respectively an $m\times n$ and a $p\times n$ dimensional real matrices. Then there exists an $m\times p$ dimensional real matrix $X$, such that the matrix $A + XB$ is of FCR, if and only if $m\geq n$ and the matrix ${\rm\bf col}\{A,\; B\}$ is of FCR.
\label{lemma:2}
\end{lemma}

The following lemma is derived in \cite{zz2020}, which is of great help in the following investigations through exploiting
the block diagonal structure of the associated matrices.

\begin{lemma}
Assume that $A_{i}^{[j]}|_{i=1,j=1}^{i=3,j=m}$ and $B_{i}^{[j]}|_{i=1,j=1}^{i=3,j=m}$ are some matrices having compatible dimensions, and the matrix $\left[A_{2}^{[1]}\;A_{2}^{[2]}\;\cdots\; A_{2}^{[m]}\right]$ is of FCR. Then the matrix
\begin{displaymath}
\left[\!\!\begin{array}{ccc}
{\rm\bf diag}\!\left\{\!A_{1}^{[1]},\;A_{2}^{[1]},\; A_{3}^{[1]}\!\right\} &  \cdots &
{\rm\bf diag}\!\left\{\!A_{1}^{[m]},\;A_{2}^{[m]},\; A_{3}^{[m]}\!\right\}   \\
\left[ B_{1}^{[1]} \;\;\;\; B_{2}^{[1]} \;\;\;\; B_{3}^{[1]}\right] &  \cdots &
\left[ B_{1}^{[m]} \;\;\;\; B_{2}^{[m]} \;\;\;\; B_{3}^{[m]}\right]\end{array}\!\!\right]
\end{displaymath}
is of FCR, if and only if the following matrix has this property
\begin{displaymath}
\left[\begin{array}{ccc}
{\rm\bf diag}\left\{A_{1}^{[1]},\; A_{3}^{[1]}\right\} &  \cdots &
{\rm\bf diag}\left\{A_{1}^{[m]},\; A_{3}^{[m]}\right\}   \\
\left[ B_{1}^{[1]} \;\;\;\;  B_{3}^{[1]}\right] &  \cdots &
\left[ B_{1}^{[m]} \;\;\;\;  B_{3}^{[m]}\right]\end{array}\right]
\end{displaymath}
\label{lemma:3}
\end{lemma}

\section{NDS Causality/Impulse Freeness}

For each ${\rm\bf \#}={\it x}$, ${\it v}$ or ${\it z}$, define a vector $\#(t)$ as $\#(t)={\rm\bf col}\left\{\#(t,i)|_{i=1}^{N}\right\}$. Moreover, define matrices $D_{\rm\bf u}$ and $E$ respectively as $D_{\rm\bf u}\!\!=\!\!{\rm\bf diag}\!\left\{D_{\rm\bf u}(i)|_{i=1}^{N}\!\right\}$ and $E\!\!=\!\!{\rm\bf diag}\!\left\{E(i)|_{i=1}^{N}\!\right\}$.  In addition, define matrices $A_{\rm\bf
*\#}$, $B_{\rm\bf *}$ and $C_{\rm\bf *}$ with ${\rm\bf *,\#}={\rm\bf x}$, ${\rm\bf y}$, ${\rm\bf v}$ or ${\rm\bf z}$ respectively as $A_{\rm\bf
*\#}=\!\!{\rm\bf diag}\!\left\{A_{\rm\bf
*\#}(i)|_{i=1}^{N}\!\right\}$, $B_{\rm\bf *}\!\!=\!\!{\rm\bf diag}\!\left\{B_{\rm\bf
*}(i)|_{i=1}^{N}\!\right\}$, $C_{\rm\bf *}\!=\!{\rm\bf diag}\!\left\{C_{\rm\bf
*}(i)|_{i=1}^{N}\!\right\}$. With these symbols, the dynamics of all the subsystems in the NDS $\rm\bf\Sigma$ can be compactly represented by
\begin{equation}
\left[\! \begin{array}{c}
{E\delta(x(t))}\\
{{{z}}(t)}\\
{{y}(t)}
\end{array} \!\right] = \left[\! \begin{array}{ccc}
{A_{\rm\bf xx}} & {A_{\rm\bf xv}} & {B_{\rm\bf x}}\\
{A_{\rm\bf zx}} & {A_{\rm\bf zv}} & {B_{\rm\bf z}}\\
{C_{\rm\bf x}} & {C_{\rm\bf v}} & {D_{\rm\bf u}}
\end{array} \!\right]\left[\! \begin{array}{c}
{x(t)}\\
{{v}(t)}\\
{u(t)}
\end{array} \!\right]
\label{eqn:8}
\end{equation}

Assume that the NDS $\rm\bf\Sigma$ is well-posed. Satisfaction of this assumption is essential for the NDS $\rm\bf\Sigma$ to work properly and is equivalent to that the matrix $I - A_{\rm\bf zv}\Phi$ is invertible \cite{Zhou2015,zyl2018}. Under this assumption, substitute Equation (\ref{eqn:2}) into the above equation. Then a descriptor model can be obtained for the dynamics of the NDS $\rm\bf\Sigma$ that has completely the same form as that of Equation (\ref{eqn:3}). Particularly, the matrices $A$, $B$, $C$ and $D$ are given by the following linear fractional transformation of the SCM $\Phi$,
\begin{eqnarray}
\hspace*{-0.5cm}\left[\!\!\begin{array}{cc} A & B \\  C & D
\end{array}\!\!\right] \!\!\!\!&=&\!\!\!\!\left[\!\!\begin{array}{cc}
A_{\rm\bf xx} & \hspace*{-0.2cm} B_{\rm\bf x} \\
C_{\rm\bf x}  & \hspace*{-0.2cm} D_{\rm\bf u} \end{array}\!\!\right]+\left[\!\!\begin{array}{c}
A_{\rm\bf xv} \\
C_{\rm\bf v}\end{array}\!\!\right]\times  \nonumber\\
& & \hspace*{0.2cm} \Phi\!
\left[\;I\!-\!A_{\rm\bf zv}\Phi\;\right]^{\!-1}\!\left[
A_{\rm\bf zx}\;\; B_{\rm\bf z}\right]
\label{eqn:9}
\end{eqnarray}

From this lumped expression, it is clear that the models adopted in \cite{zlr2015,wl2018} are its special cases. In addition, the model given by Equations (\ref{eqn:1}) and (\ref{eqn:2}) is more efficient in representing structure characteristics of a large scale NDS, noting that the inverse of a sparse matrix is in general not sparse \cite{Gantmacher1959,hj1991,Zhou2015,Zhou2020}.

While the above expressions make the conclusions of Lemma \ref{lemma:1} in principle applicable to the verification of the causality/impulse-mode non-existence of the NDS $\rm\bf\Sigma$, a direct application may meet serious numerical difficulties, especially when the NDS under investigation is constituted from a large number of subsystems, noting that matrix inversions are involved in Equation (\ref{eqn:9}), which is usually computationally complicated and numerically unstable with a high dimensional matrix. In addition, calculating the rank of a high dimensional matrix is also computationally challenging, as well as checking whether or not a high dimensional matrix is of FCR \cite{Gantmacher1959, gv1996, hj1991}.

On the other hand, from Lemma \ref{lemma:1}, it is clear that if there are some efficient methods to verify separately the condition ${\rm\bf rank} \left([E\;\; B]\right)=m$ and the condition that the matrix $U_{E2}^{T}AV_{E2}$ is of FCR, then no matter the descriptor system is of continuous time or of discrete time, the associated causality and non-existence of an impulse mode can be efficiently determined.

As mentioned in the previous section, causality/impulse freeness verification becomes trivial when the matrix $E$ is of FCR. Hence, this paper only investigates the situation in which the matrix $E$ is not of FCR.

On the basis of Lemma \ref{lemma:1}, as well as properties of a linear fractional transformation, a computationally efficient condition is derived respectively for the satisfaction of Conditions I and II by the NDS $\rm\bf\Sigma$. To simplify expressions, $n_{\rm\bf \star}$ with $\star=x$,  $z$, $v$, $u$ and $y$, are adopted to represent respectively the dimensions of the vectors $x(t)$, $z(t)$, $v(t)$, $u(t)$ and $y(t)$ of the NDS $\rm\bf\Sigma$. Obviously, $n_{\rm\bf \star}=\sum_{i=1}^{N}n_{\rm\bf \star}(i)$. In addition, $n_{\rm\bf e}$ is adopted to denote $\sum_{i=1}^{N}n_{\rm\bf e}(i)$.

\begin{theorem}
Let matrices $U_{E2}$ and $V_{E2}$ have the same definitions as those of Lemma \ref{lemma:1} with $n$ being replaced by $n_{\rm\bf x}$. If the matrix $U_{E2}^{T}\left[ A_{\rm\bf xx}V_{E2}\;\;\;  A_{\rm\bf xv}\right]$ is of FCR, then Condition I is always satisfied by the NDS $\rm\bf\Sigma$, no matter how the subsystems are connected. Otherwise, partition the matrix $\left(U_{E2}^{T}\left[ A_{\rm\bf xx}V_{E2}\;\;\;  A_{\rm\bf xv}\right]\right)^{\perp}$ as
\begin{equation}
\left(U_{E2}^{T}\left[ A_{\rm\bf xx}V_{E2}\;\;\;  A_{\rm\bf xv}\right]\right)^{\perp}
=\left[\begin{array}{c}
N_{\rm\bf xx} \\ N_{\rm\bf xv} \end{array}\right]
\label{eqn:10}
\end{equation}
in which the matrix $N_{\rm\bf xx}$ has $n_{\rm\bf x}-r$ rows. Then the NDS $\rm\bf\Sigma$ satisfies Condition I, if and only if the matrix $N_{\rm\bf xv} - \Phi\left(A_{\rm\bf zx} V_{E2} N_{\rm\bf xx} + A_{\rm\bf zv} N_{\rm\bf xv}\right)$ is of FCR.
\label{theorem:1}
\end{theorem}

The proof of the above theorem is deferred to the appendix.

Clearly, matrix inversions of Equation (\ref{eqn:9}) are no longer required in the above condition. This is quite attractive in large scale NDS  analysis and synthesis. In addition, the matrix $N_{\rm\bf xv} - \Phi\left(A_{\rm\bf zx} V_{E2} N_{\rm\bf xx} + A_{\rm\bf zv} N_{\rm\bf xv}\right)$ depends affinely on the SCM $\Phi$, which is helpful in NDS parameter selections and topology designs, as well as subsystem dynamics selections, recalling that subsystem parameters can also be included in this matrix \cite{Zhou2020,zz2020}. Specifically, requirements on subsystem dynamics are made clear in the next section using this result, from which a causal/impulse free NDS can be constructed.

It is worthwhile to mention that a large scale NDS usually has a sparse structure \cite{sbkkmpr2011,Siljak1978,zyl2018}. This implies that the dimension of the matrix $N_{\rm\bf xv} - \Phi\left(A_{\rm\bf zx} V_{E2} N_{\rm\bf xx} + A_{\rm\bf zv} N_{\rm\bf xv}\right)$ is usually significantly smaller than that of the state vector $x(t)$ of the NDS $\rm\bf\Sigma$. That is, compared with the dimension of the matrix $E$ in Equation (\ref{eqn:8}) and that of the matrix $A$ in Equation (\ref{eqn:9}), which are respectively $n_{\rm\bf e}\times n_{\rm\bf x}$ and $n_{\rm\bf x}\times n_{\rm\bf x}$, this matrix often has a much lower dimension. This is also very attractive from the computational viewpoint.

Note that the matrices $E$, $A_{\rm\bf xx}$, $A_{\rm\bf xv}$, $A_{\rm\bf zx}$ and $A_{\rm\bf zv}$ have a consistent block diagonal structure. This means that the SVD of the matrix $E$, as well as the computation of the matrices  $\left(U_{E2}^{T}\left[ A_{\rm\bf xx}V_{E2}\;\;\;  A_{\rm\bf xv}\right]\right)^{\perp}$ and $A_{\rm\bf zx} V_{E2} N_{\rm\bf xx} + A_{\rm\bf zv} N_{\rm\bf xv}$, can be calculated with each individual subsystem separately. Moreover, the matrices $N_{\rm\bf xv}$ and $A_{\rm\bf zx} V_{E2} N_{\rm\bf xx} + A_{\rm\bf zv} N_{\rm\bf xv}$ are also block diagonal. These characteristics are completely the same as those of the conditions established in \cite{Zhou2015,Zhou2020,zz2020} for NDS controllability/observability. This means that the condition of Theorem \ref{theorem:1} shares the same computational advantages with the conditions in these works. Specifically, computation costs for the associated matrices, that is, the matrices $N_{\rm\bf xv}$ and $A_{\rm\bf zx} V_{E2} N_{\rm\bf xx} + A_{\rm\bf zv} N_{\rm\bf xv}$, increase linearly with the subsystem number $N$, while those using a lumped descriptor model, that is, the matrices $A$ and $U_{E2}^{T}AV_{E2}$, increase in an order of at least $N^{3}$. In addition, the associated computations are numerically more stable. A detailed analysis can be found in \cite{Zhou2015}, together with some comparisons through numerical examples.

In order to guarantee that the matrix $U_{E2}^{T}\left[ A_{\rm\bf xx}V_{E2}\;\;\;  A_{\rm\bf xv}\right]$ is of FCR, it is obvious that the matrix $U_{E2}^{T}A_{\rm\bf xx}V_{E2}$ must have a FCR. From Theorem \ref{theorem:1} and the consistent block diagonal structure of the matrices $E$ and $A_{\rm\bf xx}$, the latter is equivalent to that each subsystem ${\rm\bf\Sigma}_{i}$ with $i\in \{1,\;2,\;\cdots,\;N\}$ holds this property when it is isolated from other subsystems. That is, for each $i=1,2,\cdots,N$, the matrix pair $(E(i),\; A_{\rm\bf xx}(i))$ satisfies Condition I. On the other hand, Theorem \ref{theorem:1} reveals that satisfaction of Condition I by each individual subsystem is not necessary for the satisfaction of the whole NDS $\rm\bf\Sigma$. That is, even if there are some subsystems that are not causal or have an impulse mode by themselves, it is still possible to construct an NDS $\rm\bf\Sigma$ though topology designs that is causal/free from impulse modes.

Define a set ${\cal S}_{I}$ as
\begin{displaymath}
{\cal S}_{I}=\left\{i\left| \begin{array}{c} U_{E2}^{T}(i)\left[ A_{\rm\bf xx}(i)V_{E2}(i)\;\;\;  A_{\rm\bf xv}(i)\right] {\rm \;\; is\;\; of\;\; FCR,}\\ 1\leq i\leq N \end{array}  \right.\right\}
\end{displaymath}
Assume that the matrix $U_{E2}^{T}\left[ A_{\rm\bf xx}V_{E2}\;\;\;  A_{\rm\bf xv}\right]$ is not of FCR, but the set ${\cal S}_{I}$ is not empty. Then from Lemma \ref{lemma:3} and Equation (\ref{eqn:a2}), as well as the consistent block diagonal structure of the associated matrices, it can be declared that removing all the columns in the matrix $\Pi$ associated with a subsystem ${\rm\bf\Sigma}_{i}$ with $i\in {\cal S}_{I}$, does not change the conclusions about whether or not the matrix $\Pi$ is of FCR. It can therefore be declared from the proof of Theorem \ref{theorem:1} that the removal of these subsystem associated columns does not alter the conclusions about the satisfaction of Condition I. On the other hand, the removal of these columns in the matrix $\Pi$ will lead to a necessary and sufficient condition with a lower computation costs. The larger the number of the elements in the set ${\cal S}_{I}$ is, the more the computation costs are reduced. The details are very similar to NDS controllability/observability verification discussed in \cite{zz2020,Zhou2020}, and are therefore omitted.

Through investigating properties of the left null space of the matrix $[E\;\; B]$, similar results can be established for verifying the condition that the rank of this matrix is equal to $n_{\rm\bf e}$.

\begin{theorem}
Assume that $n_{\rm\bf x}=n_{\rm\bf e}$. Let matrix $U_{E2}$ has the same definition as that of Theorem \ref{theorem:1}. If the matrix ${\rm\bf col}\left\{U_{E2}^{T} B_{\rm\bf x},\; B_{\rm\bf z}\right\}$ is of full row rank (FRR), then the NDS $\rm\bf\Sigma$ always satisfies Condition II with $m$ being replaced by $n_{\rm\bf e}$. Otherwise, let $\left[N_{\rm\bf Bx} \;\; N_{\rm\bf Bz} \right]$ be a matrix whose rows form a base of the left null space of the aforementioned matrix, in which the submatrix $N_{\rm\bf Bx}$ has $n_{\rm\bf e}-r$ columns. Then the NDS $\rm\bf\Sigma$ satisfies ${\rm\bf rank} \left([E\;\; B]\right)=n_{\rm\bf e}$, if and only if the matrix $N_{\rm\bf Bz} - \left(N_{\rm\bf Bx} U_{E2}^{T} A_{\rm\bf xv} + N_{\rm\bf Bz} A_{\rm\bf zv}\right)\Phi$ is of FRR.
\label{theorem:2}
\end{theorem}

The proof of the above theorem is provided in the appendix.

From the consistent block diagonal structure of the associated matrices, it is clear that the condition of Theorem \ref{theorem:2} has the same computational advantages as those of Theorem \ref{theorem:1} in NDS analysis and synthesis.

Define a set ${\cal S}_{II}$ as
\begin{displaymath}
{\cal S}_{II}=\left\{i\left| \begin{array}{c} {\rm\bf col}\left\{U_{E2}^{T}(i) B_{\rm\bf x}(i),\; B_{\rm\bf z}(i)\right\} {\rm \;\; is\;\; of\;\; FRR,}\\ 1\leq i\leq N \end{array}  \right.\right\}
\end{displaymath}
When this set is not empty, similar arguments as those for Theorem 1 show that the removal of all these subsystem associated rows in the counterpart matrix does not alter the conclusions about the satisfaction of Condition II, but this removal can reduce computation costs in Condition II verifications.

Note that the matrix ${\rm\bf col}\left\{U_{E2}^{T} B_{\rm\bf x},\; B_{\rm\bf z}\right\}$ is of FRR only when the matrix $U_{E2}^{T} B_{\rm\bf x}$ holds this property. From the consistent block diagonal structure of the matrices $E$ and $B_{\rm\bf x}$, as well as the proof of Theorem \ref{theorem:2}, the latter is equivalent to that for each $i=1,2,\cdots,N$, ${\rm\bf rank} \left([E(i)\;\; B_{\rm\bf x}(i)]\right)=n_{\rm\bf e}(i)$, in which $n_{\rm\bf e}(i)$ stands for the number of the rows of the matrix $E(i)$ in the $i$-th subsystem ${\rm\bf\Sigma}_{i}$. In addition, Theorem \ref{theorem:2} also makes it clear that this condition is not necessary for the whole NDS $\rm\bf\Sigma$ to meet the requirement ${\rm\bf rank} \left([E\;\; B]\right)=n_{\rm\bf e}$. Once again, this implies that even if there are some subsystems that are not causal, it is still possible to build a causal NDS $\rm\bf\Sigma$ though selecting appropriate subsystem connections.

For each $i=1,2,\cdots,N$, let
\begin{displaymath}
E(i)\!=\! U_{E}(i) \!\!\left[\!\!\!\begin{array}{cc}{\rm\bf diag}\{\!\sigma_{Ej}(i)|_{j=1}^{r(i)}\!\} & 0_{r(i)\times [n_{\rm\bf x}(i)-r(i)]} \\ 0_{[n_{\rm\bf e}(i)-r(i)]\times r(i)} & 0_{[n_{\rm\bf e}(i)\!-\!r(i)]\times [n_{\rm\bf x}(i)\!-\!r(i)]} \end{array}\!\!\!\right] \!\! V_{E}^{T}(i)
\end{displaymath}
be the SVD of the matrix $E(i)$. Moreover, divide the matrices $U_{E}(i)$ and $V_{E}(i)$ respectively as
\begin{displaymath}
U_{E}(i)=\left[U_{E1}(i)\; U_{E2}(i)\right], \hspace{0.5cm}
V_{E}(i)=\left[V_{E1}(i)\; V_{E2}(i)\right]
\end{displaymath}
in which $U_{E1}(i)\in {\cal R}^{n_{\rm\bf e}(i)\times r(i)}$, $U_{E2}(i)\in {\cal R}^{n_{\rm\bf e}(i)\times [n_{\rm\bf e}(i)-r(i)]}$, $V_{E1}(i)\in {\cal R}^{n_{\rm\bf x}(i)\times r(i)}$ and $V_{E2}(i)\in {\cal R}^{n_{\rm\bf x}(i)\times [n_{\rm\bf x}(i)-r(i)]}$. When the matrix $E(i)$ is of FRR that is equivalent to $r(i)=n_{\rm\bf e}(i)$, the matrix $U_{E2}(i)$ vanishes. On the other hand, if the matrix $E(i)$ is of FCR that leads to $r(i)=n_{\rm\bf x}(i)$, then the matrix $V_{E2}(i)$ vanishes. From Theorems \ref{theorem:1} and \ref{theorem:2}, as well as the consistent block diagonal structure of the associated matrices, the following results can be straightforwardly obtained. Their proof are omitted due to its obviousness.

\begin{corollary}
Assume that the dynamics of an NDS is described by Equations (\ref{eqn:1}) and (\ref{eqn:2}).
\begin{itemize}
\item If for each $i\in\{1,\;2,\;\cdots,\;N\}$, the matrix $E(i)$ is of FRR or the matrix $U_{E2}^{T}(i)\left[ A_{\rm\bf xx}(i)V_{E2}(i)\;\;\;  A_{\rm\bf xv}(i)\right]$ is of FCR, then the NDS $\rm\bf\Sigma$ always satisfies Condition I, no matter how its subsystems are connected.
\item If for an arbitrary  $i\in \{1,2,\cdots,N\}$, the matrix ${\rm\bf col}\left\{U_{E2}^{T}(i) B_{\rm\bf x}(i),\; B_{\rm\bf z}(i)\right\}$ is of FRR, then the NDS $\rm\bf\Sigma$ always satisfies Condition II, no matter how its subsystems are connected.
\end{itemize}
\label{corollary:0}
\end{corollary}

Note that the above conditions are imposed on each individual subsystem independently. This means that there exist some kinds of plants which always form a causal/impulse free NDS, no matter how these plants are connected to each other. These characteristics are quite important in some real world problems. For example, in a system consisting of several autonomous agents, connections among these agents may change according to working situations, and/or may be difficult to predict in practice \cite{sbkkmpr2011,zyl2018}.

It is worthwhile to point out that the matrix $U_{E2}^{T}(i)\left[ A_{\rm\bf xx}(i)V_{E2}(i)\;\;\;  A_{\rm\bf xv}(i)\right]$ is of FCR, only when the matrix $U_{E2}^{T}(i)A_{\rm\bf xx}(i)V_{E2}(i)$ is of FCR; while the matrix ${\rm\bf col}\left\{U_{E2}^{T}(i) B_{\rm\bf x}(i),\; B_{\rm\bf z}(i)\right\}$ is of FRR, only the matrix $U_{E2}^{T}(i) B_{\rm\bf x}(i)$ is of FRR. On the other hand, it is clear from Lemma 1 that in order to guarantee that the NDS of Equations (\ref{eqn:1}) and (\ref{eqn:2}) is causal/impulse free, it is necessary that $n_{\rm\bf e}\geq n_{\rm\bf x}$, that is, $\sum_{i=1}^{N} n_{\rm\bf e}(i) \geq  \sum_{i=1}^{N} n_{\rm\bf x}(i)$. However, this is not equivalent to $n_{\rm\bf e}(i) \geq n_{\rm\bf x}(i)$ for each $i=1,2,\cdots N$.

\section{Decentralized Causality/Impulse Constructibility}

In the previous section, some necessary/sufficient conditions have been derived for the causality/impulse freeness of an NDS that depends affinely on its subsystem interactions. These results are meaningful in analyzing a constructed NDS. But except the selection of the SCM $\Phi$, they can not yet be directly utilized in building a causal/impulse free NDS. In this section, requirements are investigated for a subsystem on the basis of these conditions, from which a causal/impulse free NDS can be built. As in the previous section, we once again study separately the satisfaction of Conditions I and II.

From Corollary \ref{corollary:0}, it is clear that Conditions I is always satisfied when the matrix $U_{E2}^{T}\left[ A_{\rm\bf xx}V_{E2}\;\;\;  A_{\rm\bf xv}\right]$ is of FCR, while Conditions II is always satisfied when the matrix ${\rm\bf col}\left\{U_{E2}^{T} B_{\rm\bf x},\; B_{\rm\bf z}\right\}$ is of FRR. That is, these two situations are trivial for studying the aforementioned subsystem requirement problem. The following two assumptions are therefore adopted throughout the remaining of this section.
\begin{itemize}
\item the matrix $U_{E2}^{T}\left[ A_{\rm\bf xx}V_{E2}\;\;\;  A_{\rm\bf xv}\right]$ is not of FCR
\item the matrix ${\rm\bf col}\left\{U_{E2}^{T} B_{\rm\bf x},\; B_{\rm\bf z}\right\}$ is not of FRR.
\end{itemize}

The following properties of the sets ${\cal S}_{I}$ and ${\cal S}_{II}$, which are defined in the previous section, are very helpful in deriving conditions for the constructibility of a causal/impulse free NDS from descriptor form subsystem.

\begin{lemma}
Assume that both the set ${\cal S}_{I}$ and the set ${\cal S}_{II}$ are not empty.
\begin{itemize}
\item If $i\in {\cal S}_{I}$, then the matrix ${\rm\bf col}\{U_{E2}^{T}(i) A_{\rm\bf xx}(i),\; V_{E1}^{T}(i)\}$ is of FCR.
\item If $i\in {\cal S}_{I}$, then the matrix $\left[U_{E1}(i)\;\; B_{\rm\bf x}(i)\right]$ is of FRR.
\end{itemize}
\label{lemma:4}
\end{lemma}

\hspace*{-0.45cm}{\rm\bf Proof:}
From the definition of the set ${\cal S}_{I}$, we have that if $i\in {\cal S}_{I}$, then the matrix $U_{E2}^{T}(i)\left[ A_{\rm\bf xx}(i)V_{E2}(i)\;\;\;  A_{\rm\bf xv}(i)\right]$ is of FCR, which implies that the matrix $U_{E2}^{T}(i)A_{\rm\bf xx}(i)V_{E2}(i)$ is also of FCR. The latter is equivalent to
\begin{equation}
{\rm\bf null}\left( U_{E2}^{T}(i)A_{\rm\bf xx}(i) \right)
\bigcap
{\rm\bf span}\left( V_{E2}(i) \right) =\{ 0 \}
\label{eqn:4}
\end{equation}

Recall that $V_{E}(i)=\left[ V_{E1}(i)\;\; V_{E2}(i) \right]$ is an orthogonal matrix. It is straightforward to show that ${\rm\bf span}\left( V_{E2}(i) \right) = {\rm\bf null}\left( V_{E1}^{T}(i) \right)$. Substitute this relation into Equation (\ref{eqn:4}), we have that ${\rm\bf null}\left( U_{E2}^{T}(i)A_{\rm\bf xx}(i) \right)
\bigcap
{\rm\bf null}\left( V_{E1}^{T}(i) \right) =\{ 0 \}$, which is equivalent to that the matrix ${\rm\bf col}\{U_{E2}^{T}(i) A_{\rm\bf xx}(i),\; V_{E1}^{T}(i)\}$ is of FCR.

On the basis that the matrix $U_{E}(i)=\left[ U_{E1}(i)\;\; U_{E2}(i) \right]$ is an orthogonal matrix, the second conclusion can be proven similarly. The details are omitted due to their close similarities.

This completes the proof.   \hspace{\fill}$\Diamond$

From these results, some necessary and sufficient conditions are established for the existence of a SCM $\Phi$, such that the associated NDS $\rm\bf\Sigma$ of Equations (\ref{eqn:1}) and (\ref{eqn:2}) is causal/impulse free.

\begin{theorem}
There exists a SCM $\Phi$ such that Condition I is satisfied by the NDS $\rm\bf\Sigma$, if and only if for each $i\not\in {\cal S}_{I}$, the matrix ${\rm\bf col}\left\{ U_{E2}^{T}(i)A_{\rm\bf xx}(i),\; A_{\rm\bf zx}(i),\;
V_{E1}^{T}(i) \right\}$ is of FCR.
\label{theorem:3}
\end{theorem}

Similarly, the following results can be derived for Condition II.

\begin{theorem}
There exists a SCM $\Phi$ such that Condition II with $m$ being replaced by $n_{\rm\bf e}$ is satisfied by the NDS $\rm\bf\Sigma$, if and only if for each $i\not\in {\cal S}_{II}$, the matrix $\left[ U_{E1}(i),\; A_{\rm\bf xv}(i),\; B_{\rm\bf x}(i) \right]$ is of FRR.
\label{theorem:4}
\end{theorem}

It is worthwhile to emphasize that for a discrete time descriptor system to be causal, Conditions I and II must be simultaneously satisfied, together with a dimension requirement and a regularity requirement. On the other hand, simultaneous satisfaction of the conditions of Theorems \ref{theorem:3} and \ref{theorem:4} does not imply the existence of a SCM $\Phi$ that meets simultaneously these two conditions. In other words, these two theorems generally only give a necessary condition on a discrete time descriptor form subsystem, from which a causal NDS can be constructed. This is different from that about a continuous time descriptor form subsystem, for which Theorem \ref{theorem:3} provides a necessary and sufficient condition for constructing an impulse free NDS, provided that the dimension condition $n_{\rm\bf e}\geq n_{\rm\bf x}$ is satisfied which can be simply and directly verified. On the other hand, the previous section and \cite{Zhou2020} have made it clear that there are some situations under which the regularity of an NDS and its satisfaction of Condition I/II do not depend on the SCM $\Phi$. Under these situations, the conditions of Theorem \ref{theorem:3} or Theorem\ref{theorem:4} become both necessary and sufficient for the existence of a SCM $\Phi$ that makes the discrete time NDS causal.

These remarks remain valid for the following Corollaries \ref{corollary:1} and \ref{corollary:2}.

When a descriptor system is not causal/impulse free, it is usually hoped to make it reach this property through introducing some static feedbacks \cite{Dai1989a,Duan2010,zlr2015}. When an NDS is constituted from a large number of descriptor subsystems, lumped feedbacks usually lead to prohibitive hardware and communication costs, etc., and are therefore not widely appreciated. A much more feasible and extensively appreciative approach is to use local feedbacks, which is usually called decentralized control \cite{sbkkmpr2011,Siljak1978,wl2018,zlr2015}. To investigate this possibility for a descriptor form NDS, the following concept is introduced.

\begin{definition}
The NDS of Equations (\ref{eqn:1}) and (\ref{eqn:2}) are called causality/impulse constructible through a decentralized static output feedback, if and only if there exist a SCM $\Phi$ and a local control law with the following form
\begin{equation}
u(t,i)=F(i)\left[ y(t,i)+\gamma(t,i) \right]
\label{eqn:13}
\end{equation}
in which $i=1,2,\cdots,N$, such that the overall closed-loop system is causal/impulse free.
\label{def:2}
\end{definition}

Obviously, this definition includes local static state feedback as a special case, noting that the above feedback vanishes to the static state feedback $u(t,i)=F(i)\left[x(t,i)+\gamma(t,i)\right]$, provided that the matrices $C_{\rm\bf x}(i)$, $C_{\rm\bf v}(i)$ and $D_{\rm\bf u}(i)$ are respectively set as $C_{\rm\bf x}(i)=I_{n_{\rm\bf x}(i)}$, $C_{\rm\bf v}(i)=0_{n_{\rm\bf x}(i)\times n_{\rm\bf v}(i)}$ and $D_{\rm\bf u}(i)=0_{n_{\rm\bf x}(i)\times n_{\rm\bf u}(i)}$.

In descriptor system analysis and synthesis, a well established concept is causality/impulse controllability \cite{hm1999,it2001,zlr2015}. Specifically, a descriptor system is called causality/impulse controllable, if there exists a feedback such that the closed-loop system is causal/impulse free. When an NDS is under investigation, both subsystem interactions and local feedbacks can be utilized to make it causal/impulse free. While subsystem interactions can be regarded as a special kind of feedbacks, their capabilities are restricted by subsystem dynamics. It therefore appears more appropriate to use a different concept to clarify these differences.

Define a matrix $F$ as  $F\!\!=\!\!{\rm\bf diag}\!\left\{F(i)|_{i=1}^{N}\!\right\}$. Then the feedback law of Equation (\ref{eqn:13}) can be compactly rewritten as
\begin{equation}
u(t)=F \left[ y(t)+\gamma(t) \right]
\label{eqn:16}
\end{equation}
in which $\gamma(t)\!\!=\!\!{\rm\bf col}\!\left\{\gamma(t,i)|_{i=1}^{N}\!\right\}$. Substitute this relation into Equation (\ref{eqn:8}), direct algebraic manipulations show that
\begin{equation}
u(t)=(I_{n_{\rm\bf u}}-FD_{\rm\bf u})^{-1}F \left[ C_{\rm\bf x}x(t)+ C_{\rm\bf v}v(t) + \gamma(t) \right]
\label{eqn:16a}
\end{equation}
Moreover, Equation (\ref{eqn:8}) can be rewritten as
\begin{eqnarray}
\left[\! \begin{array}{c}
{E\delta(x(t))}\\
{{{z}}(t)}\\
{{y}(t)}
\end{array} \!\right] \!\!\!\!&=&\!\!\!\!\left\{ \left[\! \begin{array}{ccc}
{A_{\rm\bf xx}} & {A_{\rm\bf xv}} & {0_{n_{\rm\bf e}\times n_{\rm\bf y}}}\\
{A_{\rm\bf zx}} & {A_{\rm\bf zv}} & {0_{n_{\rm\bf z}\times n_{\rm\bf y}}}\\
{C_{\rm\bf x}} & {C_{\rm\bf v}} & {0_{n_{\rm\bf y}\times n_{\rm\bf y}}}
\end{array} \!\right] +
\left[\! \begin{array}{c}
{B_{\rm\bf x}}\\
{B_{\rm\bf z}}\\
{D_{\rm\bf u}}  \end{array} \!\right]\times \right. \nonumber \\
& &\hspace*{-2cm}\left.\left(I_{n_{\rm\bf u}}-FD_{\rm\bf u}\right)^{-1}F\!\!\!\!
\left.\begin{array}{c} \\
{\left[\!\!\begin{array}{ccc} C_{\rm\bf x}  & C_{\rm\bf v} & I_{n_{\rm\bf y}} \end{array}\!\!\right]} \\
\\
\end{array}\right.\!\!\!\! \right\}
\left[\! \begin{array}{c}
{x(t)}\\
{{v}(t)}\\
{\gamma(t)}
\end{array} \!\right]
\label{eqn:17}
\end{eqnarray}

To investigate decentralized causal/impulse constructibility, define Sets ${\cal D}_{I}$ and ${\cal D}_{II}$ respectively as
\begin{eqnarray*}
& & {\cal D}_{I}=\left\{i\left| \begin{array}{r} {\rm\bf col}\left\{ U_{E2}^{T}(i)A_{\rm\bf xx}(i),\; A_{\rm\bf zx}(i),\;
V_{E1}^{T}(i) \right\} \\  {\rm \;\; is\;\; not \;\; of\;\; FCR,}\;\;\;1\leq i\leq N \end{array}  \right.\right\}  \\
& & {\cal D}_{II}=\left\{i\left| \begin{array}{r} \left[ U_{E1}(i),\; A_{\rm\bf xv}(i),\; B_{\rm\bf x}(i) \right] {\rm \;\; is\;\;not \;\; of} \\ {\rm\;\; FRR,}\;\;\; 1\leq i\leq N \end{array}  \right.\right\}
\end{eqnarray*}

On the basis of the relation of Equation (\ref{eqn:17}), the following conclusions can be derived for the decentralized causality/impulse constructibility of the NDS $\rm\bf\Sigma$.

\begin{corollary}
Assume that the set ${\cal D}_{I}$ is not empty. Then there exist a decentralized static output feedback of Equation (\ref{eqn:13}) and a SCM $\Phi$, such that the corresponding NDS $\rm\bf\Sigma$ described by Equations (\ref{eqn:1}) and (\ref{eqn:2}) satisfies Condition I, if and only if for each $i\in{\cal D}_{I}$, the matrix ${\rm\bf col}\left\{ U_{E2}^{T}(i)A_{\rm\bf xx}(i),\; A_{\rm\bf zx}(i),\; C_{\rm\bf x}(i),\;
V_{E1}^{T}(i) \right\}$ is of FCR.
\label{corollary:1}
\end{corollary}

Note that when the matrix ${\rm\bf col}\left\{ U_{E2}^{T}(i)A_{\rm\bf xx}(i),\; A_{\rm\bf zx}(i),\;
V_{E1}^{T}(i) \right\}$ is of FCR, it is certain that the matrix ${\rm\bf col}\left\{ U_{E2}^{T}(i)A_{\rm\bf xx}(i),\; A_{\rm\bf zx}(i),\; C_{\rm\bf x}(i),\;
V_{E1}^{T}(i) \right\}$ is of FCR, also. The contrary, however, is in general not true. This means that local static output feedbacks are usually helpful in transforming a continuous time NDS from having an impulse mode to impulse free. In addition, this kind of feedbacks are in general also helpful in transforming a discrete time NDS from non-causal to causal, provided its non-causality is resulted from the violation of Condition I.

The following results, however, reveal that when Condition II is violated, causality of a discrete time NDS can not be achieved through this kind of feedbacks.

\begin{corollary}
Assume that the set ${\cal D}_{II}$ is not empty. Then there does not exist a decentralized static output feedback of Equation (\ref{eqn:13}), such that the corresponding NDS $\rm\bf\Sigma$ described by Equations (\ref{eqn:1}) and (\ref{eqn:2}) satisfies Condition II through appropriately selecting a SCM $\Phi$.
\label{corollary:2}
\end{corollary}

The above corollary makes it clear that if there is a subsystem ${\rm\bf\Sigma}_{i}$ whose system matrices do not meet the requirement that
$\left[ U_{E1}(i),\; A_{\rm\bf xv}(i),\; B_{\rm\bf x}(i) \right]$ is of  FRR, then in order to construct a causal NDS, the only option is to modify its system matrices $E(i)$ and/or $A_{\rm\bf xv}(i)$ and/or $B_{\rm\bf x}(i)$ through adjusting its dynamics.

\section{Concluding Remarks}

Through analyzing the structure of the null space of the associated matrix, a novel matrix rank based necessary and sufficient condition is derived in this paper for the causality/impulse mode freeness of a descriptor system that is constituted from several descriptor form subsystems. This condition successfully avoid utilizing a lumped model of the networked dynamic system in which matrix inversions are required. This condition also clarifies situations in which NDS causality/impulse mode freeness is completely and independently determined by the dynamics of each individual subsystem. In addition, the associated matrix in this condition depends affinely on subsystem connections, which is helpful in system parameter selections and topology designs. A prominent characteristic of this condition is that all the involved numerical computations are performed independently on each individual subsystem. This property is quite attractive for large scale NDS analysis and synthesis, in which both computation costs and numerical stability are essential issues.

In addition, requirements have also been made clear on each descriptor form subsystem, such that a causal/impulse free NDS can be constructed from them. It has been shown that a local static output feedback is in general helpful in transforming an NDS from having an impulse mode to impulse free, but there exist some situations under which this feedback is not effective in converting a non-causal NDS into causal.

Some further efforts, however, are still required to find a necessary and sufficient condition for the existence of a SCM $\Phi$, such that the associated NDS satisfies simultaneously Conditions I and II. As a further topic, it is interesting to see possibilities of developing a  method that constructs a causal/impulse free NDS through appropriately selecting subsystem interactions and local output feedbacks, as well as developing a numerically more efficient method for the verification of the conditions of Theorems \ref{theorem:1} and \ref{theorem:2} that explicitly utilizes the block diagonal structure of the associated matrices.

\renewcommand{\theequation}{a.\arabic{equation}}
\setcounter{equation}{0}

\section*{Appendix: Proof of Some Technical Results}

\vspace{0.5cm}
\hspace*{-0.45cm}{\rm\bf Proof of Theorem \ref{theorem:1}:}
For brevity, define matrices $\Omega$ and $\Pi$ respectively as
\begin{eqnarray}
\Omega \!\!\!\!&=&\!\!\!\! U_{E2}^{T}\left[ A_{\rm\bf xx} + A_{\rm\bf xv}\Phi\left( I-A_{\rm\bf zv} \Phi \right)^{-1} A_{\rm\bf zx} \right] V_{E2}
\label{eqn:a1}\\
\Pi \!\!\!\!&=&\!\!\!\! \left[\begin{array}{cc}
U_{E2}^{T}A_{\rm\bf xx}V_{E2} &   U_{E2}^{T}A_{\rm\bf xv} \\
-\Phi A_{\rm\bf zx}V_{E2} & I-\Phi A_{\rm\bf zv}
\end{array}\right]
\label{eqn:a2}
\end{eqnarray}
From Lemma \ref{lemma:1} and Equation (\ref{eqn:9}), we have that  Condition I is satisfied by the NDS $\rm\bf\Sigma$, if and only if the matrix $\Omega$ is of FCR. The latter is equivalent to that
\begin{equation}
\Omega \alpha =0
\label{eqn:a20}
\end{equation}
if and only if $\alpha=0$.

Let $\alpha$ be an arbitrary vector satisfying Equation (\ref{eqn:a20}). Define a vector $\beta$ as $\beta= \left( I- \Phi A_{\rm\bf zv}\right)^{-1} \Phi A_{\rm\bf zx} V_{E2} \alpha$. Noting that when the matrix $I-  A_{\rm\bf zv}\Phi$ is invertible, the matrix $I-\Phi A_{\rm\bf zv}$ is also invertible. Moreover, $\left( I- \Phi A_{\rm\bf zv}\right)^{-1} \Phi = \Phi \left( I-  A_{\rm\bf zv}\Phi\right)^{-1}$. From these observations, it can be claimed that the vector ${\rm\bf col}\left\{\alpha,\;\beta\right\}$ is a solution to the following equation,
\begin{equation}
\Pi \left[\begin{array}{c}
 \alpha   \\   \beta
\end{array}\right] = 0
\label{eqn:a21}
\end{equation}
Note that when $\alpha\neq 0$, the vector ${\rm\bf col}\left\{\alpha,\;\beta\right\}$ is also not equal to zero. This means that if the matrix $\Omega$ is not of FCR, then the matrix $\Pi$ is certainly not of FCR, also.

On the contrary, assume that the matrix $\Pi$ is not of FCR. Then there is a nonzero vector ${\rm\bf col}\left\{\alpha,\;\beta\right\}$ satisfying Equation (\ref{eqn:a21}) with the sub-vectors $\alpha$ and $\beta$ having compatible dimensions. Obviously $\beta= \left( I- \Phi A_{\rm\bf zv}\right)^{-1} \Phi A_{\rm\bf zx} V_{E2} \alpha$, which implies that $\alpha\neq 0$. In addition, this nonzero vector $\alpha$ also satisfies Equation (\ref{eqn:a20}). It can therefore be declared that if the matrix $\Pi$ is not of FCR, then the matrix $\Omega$ is certainly not invertible.

These arguments mean that the matrix $\Omega$ is of FCR, if and only if the matrix $\Pi$ is of FCR.

Obviously, if the matrix $U_{E2}^{T}\left[ A_{\rm\bf xx}V_{E2}\;\;\;  A_{\rm\bf xv}\right]$ is of FCR, then the matrix $\Pi$ is always of FCR, no matter what value the SCM $\Phi$ takes. This means that  Condition I is always satisfied by the NDS $\rm\bf\Sigma$, no matter how its subsystems are connected.

In the remaining of this proof, it is assumed that the matrix $U_{E2}^{T}\left[ A_{\rm\bf xx}V_{E2}\;\;\;  A_{\rm\bf xv}\right]$ is not of FCR, which means that the matrix $\left(U_{E2}^{T}\left[ A_{\rm\bf xx}V_{E2}\;\;\;  A_{\rm\bf xv}\right]\right)^{\perp}$ is not a zero matrix.

Assume that the matrix $\Pi$ is not of FCR. Then there exists a nonzero vector $\xi$ such that $\Pi\xi=0$. From the definition of the matrix $\Pi$, it is obvious that $U_{E2}^{T}\left[ A_{\rm\bf xx}V_{E2}\;\;\;  A_{\rm\bf xv}\right]\xi=0$. That is, this vector $\xi$ must belong to the right null space of the matrix $U_{E2}^{T}\left[ A_{\rm\bf xx}V_{E2}\;\;\;  A_{\rm\bf xv}\right]$, which means that there is a nonzero vector $\zeta$ satisfying
\begin{equation}
\xi = \left[\begin{array}{c}
N_{\rm\bf xx} \\ N_{\rm\bf xv} \end{array}\right] \zeta
\label{eqn:a22}
\end{equation}
We therefore have that
\begin{equation}
\Pi \xi = \left[\begin{array}{c}
0 \\ \left[N_{\rm\bf xv} - \Phi\left(A_{\rm\bf zx} V_{E2} N_{\rm\bf xx} + A_{\rm\bf zv} N_{\rm\bf xv}\right)\right]\zeta \end{array}\right]
\label{eqn:a23}
\end{equation}
Hence
\begin{equation}
\left[N_{\rm\bf xv} - \Phi\left(A_{\rm\bf zx} V_{E2} N_{\rm\bf xx} + A_{\rm\bf zv} N_{\rm\bf xv}\right)\right]\zeta = 0
\label{eqn:a24}
\end{equation}
As $\zeta$ is a nonzero vector, this means that the matrix $N_{\rm\bf xv} - \Phi\left(A_{\rm\bf zx} V_{E2} N_{\rm\bf xx} + A_{\rm\bf zv} N_{\rm\bf xv}\right)$ is not of FCR.

On the contrary, assume that the matrix $N_{\rm\bf xv} - \Phi\left(A_{\rm\bf zx} V_{E2} N_{\rm\bf xx} + A_{\rm\bf zv} N_{\rm\bf xv}\right)$ is not of FCR. Then there is a nonzero vector $\zeta$ satisfying $\left[N_{\rm\bf xv} - \Phi\left(A_{\rm\bf zx} V_{E2} N_{\rm\bf xx} + A_{\rm\bf zv} N_{\rm\bf xv}\right)\right]\zeta = 0$. Define a vector $\xi$ as $\xi=\left[
N_{\rm\bf xx}^{T} \;\; N_{\rm\bf xv}^{T} \right]^{T} \zeta$. Then $\xi\neq 0$ and $\Pi\xi=0$. Hence, the matrix $\Pi$ is not of FCR also.

These observations imply that when the matrix $U_{E2}^{T}\left[ A_{\rm\bf xx}V_{E2}\;  A_{\rm\bf xv}\right]$ is not of FCR, the matrix $\Pi$ is of FCR, if and only if the matrix $N_{\rm\bf xv} - \Phi\left(A_{\rm\bf zx} V_{E2} N_{\rm\bf xx} + A_{\rm\bf zv} N_{\rm\bf xv}\right)$ is.

The proof can now be completed by recalling that Condition I is satisfied by the NDS $\rm\bf\Sigma$, if and only if the matrix $\Omega$ is of FCR.   \hspace{\fill}$\Diamond$

\vspace{0.5cm}
\hspace*{-0.45cm}{\rm\bf Proof of Theorem \ref{theorem:2}:}
Obviously, the condition ${\rm\bf rank} \left([E\;\; B]\right)=n_{\rm\bf e}$ is equivalent to that the matrix $\left[E\;\; B\right]$ is of FRR.

Let $\alpha$ be an arbitrary nonzero $n_{\rm\bf x}-r$ dimensional row vector. According to the definition of the matrix $U_{E2}$, it is obvious that $\alpha U_{E2}^{T}\neq 0$. On the other hand, direct matrix multiplications show that
\begin{equation}
\alpha U_{E2}^{T}\left[E\;\; B\right] = \left[0\;\;\alpha U_{E2}^{T} B\right]
\label{eqn:a25}
\end{equation}
This implies that if the matrix $\left[E\;\; B\right]$ is of FRR, then the matrix $U_{E2}^{T}B$ is also of FRR.

On the contrary, assume that the matrix $U_{E2}^{T}B$ is of FRR. Let $\xi$ be an arbitrary nonzero row vector satisfying $\xi E=0$. From the SVD of the matrix $E$, we have that
\begin{equation}
\xi E =\xi U_{E1}{\rm\bf diag}\left\{\left.\sigma_{Ei}\right|_{i=1}^{r}\right\}V_{E1}^{T}
\label{eqn:a26}
\end{equation}
As the matrix $V_{E1}$ is of FCR and $\sigma_{Ei}>0$ for each $i=1,2,\cdots,r$, it is straightforward to prove that $\xi E=0$ if and only if there is a vector $\alpha$ such that $\xi=\alpha U_{E2}^{T}$. Moreover, $\alpha\neq 0$ whenever $\xi\neq 0$, noting that the matrix $U_{E2}$ is also of FCR. Hence
\begin{equation}
\xi\left[E\;\; B\right] = \left[0\;\;\alpha U_{E2}^{T} B\right] \neq 0
\label{eqn:a27}
\end{equation}
This means that the matrix $\left[E\;\; B\right]$ is also of FRR, noting that any nonzero vector $\xi$ satisfying $\xi E\neq 0$ certainly satisfying $\xi\left[E\;\; B\right]\neq 0$.

It can therefore be declared that the matrix $\left[E\;\; B\right]$ is of FRR, if and only if the matrix $U_{E2}^{T}B$ is.

Note that a matrix is of FRR if and only if its transpose is of FCR. On the other hand, recall from Equation (\ref{eqn:9}) that $B = B_{\rm\bf x} + A_{\rm\bf xv}\Phi\left( I-A_{\rm\bf zv} \Phi \right)^{-1} B_{\rm\bf z}$. Conclusions of this theorem can be established using similar arguments as those in the proof of Theorem \ref{theorem:2}, noting that the matrix $\left(U_{E2}^{T}B\right)^{T}$ has completely the same form as the matrix $U_{E2}^{T}AV_{E2}$ there. The details are omitted due to their close similarities.

This completes the proof.   \hspace{\fill}$\Diamond$

\vspace{0.5cm}
\hspace*{-0.45cm}{\rm\bf Proof of Theorem \ref{theorem:3}:}
When the matrix $U_{E2}^{T}\left[ A_{\rm\bf xx}V_{E2}\;\;\;  A_{\rm\bf xv}\right]$ is not of FCR, its null space contains nonzero vectors. This means that the matrix constituted from a base of its null space, that is,  the matrix ${\rm\bf col}\{N_{\rm\bf xx},\; N_{\rm\bf xv}\}$, is not a zero matrix. Note that
\begin{equation}
\left[\begin{array}{c} N_{\rm\bf xv} \\
A_{\rm\bf zx} V_{E2} N_{\rm\bf xx} + A_{\rm\bf zv} N_{\rm\bf xv}
\end{array}\right]
=
\left[\begin{array}{cc} 0 & I \\
A_{\rm\bf zx} V_{E2} &  A_{\rm\bf zv}
\end{array}\right]
\left[\begin{array}{c} N_{\rm\bf xx} \\
N_{\rm\bf xv}
\end{array}\right]
\end{equation}
It can be declared from Lemma \ref{lemma:2} that there exists a matrix $\Phi$, such that the matrix $N_{\rm\bf xv} - \Phi\left(A_{\rm\bf zx} V_{E2} N_{\rm\bf xx} + A_{\rm\bf zv} N_{\rm\bf xv}\right)$ is of FCR, if and only if the matrix $\left[\begin{array}{cc} 0 & I \\
A_{\rm\bf zx} V_{E2} &  A_{\rm\bf zv}
\end{array}\right]
\left[\begin{array}{c} N_{\rm\bf xx} \\
N_{\rm\bf xv}
\end{array}\right]$ is of FCR. Recall that the matrix ${\rm\bf col}\{N_{\rm\bf xx},\; N_{\rm\bf xv}\}$ is of FCR. The latter is equivalent to
\begin{equation}
{\rm\bf null}\left(
\left[\begin{array}{cc} 0 & I \\
A_{\rm\bf zx} V_{E2} &  A_{\rm\bf zv}
\end{array}\right]\right)
\bigcap {\rm\bf span}\left(
\left[\begin{array}{c} N_{\rm\bf xx} \\
N_{\rm\bf xv}
\end{array}\right]\right)=\left\{\; 0 \; \right\}
\label{eqn:a28}
\end{equation}

On the other hand, from the definitions of the matrices $N_{\rm\bf xx}$ and $N_{\rm\bf xv}$, it is obvious that
\begin{equation}
{\rm\bf span}\left(
\left[\begin{array}{c} N_{\rm\bf xx} \\
N_{\rm\bf xv}
\end{array}\right]\right) =
{\rm\bf null}\left(U_{E2}^{T}
\left[\begin{array}{cc} A_{\rm\bf xx} V_{E2} &  A_{\rm\bf xv}
\end{array}\right]\right)
\label{eqn:a29}
\end{equation}
This implies that the condition of Equation (\ref{eqn:a28}) can be equivalently expressed as that the following matrix is of FCR,
\begin{equation}
\left[\begin{array}{cc} 0 & I \\
A_{\rm\bf zx} V_{E2} &  A_{\rm\bf zv} \\
U_{E2}^{T}A_{\rm\bf xx} V_{E2} &  U_{E2}^{T}A_{\rm\bf xv}
\end{array}\right]
\label{eqn:a42}
\end{equation}
which is further equivalent to that the matrix $\left[\begin{array}{c}
A_{\rm\bf zx} \\
U_{E2}^{T}A_{\rm\bf xx} \end{array}\right]V_{E2}$ is of FCR.

Recall that $V_{E}=\left[V_{E1} \;\; V_{E2} \right]$ is an orthogonal matrix. It can be straightforwardly shown that ${\rm\bf span}\left(V_{E2}\right) =
{\rm\bf null}\left(V_{E1}^{T}\right)$. Through similar arguments as those of Equations (\ref{eqn:a28})-(\ref{eqn:a42}), it can be shown that the matrix $\left[\begin{array}{c}
A_{\rm\bf zx} \\
U_{E2}^{T}A_{\rm\bf xx} \end{array}\right]V_{E2}$ is of FCR, if and only if the matrix ${\rm\bf col}\left\{ A_{\rm\bf zx},\;
U_{E2}^{T}A_{\rm\bf xx},\;
V_{E1}^{T} \right\}$ has this property

From Lemma \ref{lemma:4}, we have that for each $i \in {\cal S}_{I}$, the matrix ${\rm\bf col}\left\{U_{E2}^{T}A_{\rm\bf xx}(i),\;
V_{E1}^{T}(i) \right\}$ is of FCR, which means that the matrix ${\rm\bf col}\left\{ A_{\rm\bf zx}(i),\;  U_{E2}^{T}A_{\rm\bf xx}(i),\;
V_{E1}^{T}(i) \right\}$ is also of FCR.

The proof can now be completed through Lemma \ref{lemma:3} using the consistent block diagonal structure of the matrices $A_{\rm\bf xx}$, $A_{\rm\bf zx}$, $V_{E1}$ and $U_{E2}$.   \hspace{\fill}$\Diamond$

\vspace{0.5cm}
\hspace*{-0.45cm}{\rm\bf Proof of Theorem \ref{theorem:4}:}
Note that a matrix is of FRR if and only if its transpose is FCR, and vice versa. It can therefore be declared from Lemma \ref{lemma:2} that there exists a matrix $\Phi$, such that the matrix $N_{\rm\bf Bz} - \left(N_{\rm\bf Bx} U_{E2}^{T} A_{\rm\bf xv} + N_{\rm\bf Bz} A_{\rm\bf zv}\right)\Phi$ is of FRR, if and only if the matrix $\left[\!\!\begin{array}{cc} N_{\rm\bf Bz} & N_{\rm\bf Bx} U_{E2}^{T} A_{\rm\bf xv} + N_{\rm\bf Bz} A_{\rm\bf zv} \end{array}\!\!\right]$ is of FRR.

From its definition, the matrix $\left[\begin{array}{cc} N_{\rm\bf Bx} & N_{\rm\bf Bz}  \end{array}\right]$ is obviously of FRR. Moreover,
\begin{eqnarray*}
& & \left[\begin{array}{cc} N_{\rm\bf Bz} & N_{\rm\bf Bx} U_{E2}^{T} A_{\rm\bf xv} + N_{\rm\bf Bz} A_{\rm\bf zv} \end{array}\right] \\
&=&
\left[\begin{array}{cc} N_{\rm\bf Bx} & N_{\rm\bf Bz}  \end{array}\right]
\left[\begin{array}{cc} 0  &  U_{E2}^{T} A_{\rm\bf xv} \\
 I &  A_{\rm\bf zv} \end{array}\right]
\end{eqnarray*}
Hence, the matrix $\left[\!\!\begin{array}{cc} N_{\rm\bf Bz} & N_{\rm\bf Bx} U_{E2}^{T} A_{\rm\bf xv} + N_{\rm\bf Bz} A_{\rm\bf zv} \end{array}\!\!\right]$ is of FRR, if and only if
\begin{equation}
{\rm\bf null}\left(
\left[\begin{array}{cc} 0  &  U_{E2}^{T} A_{\rm\bf xv} \\
 I &  A_{\rm\bf zv} \end{array}\right]^{T}
\right)
\bigcap {\rm\bf span}\left(
\left[\begin{array}{c} N_{\rm\bf Bx}^{T} \\
N_{\rm\bf Bz}^{T}
\end{array}\right]\right)=\left\{\; 0 \; \right\}
\label{eqn:a30}
\end{equation}

Recall that the row vectors of the matrix $\left[ N_{\rm\bf Bx} \;\; N_{\rm\bf Bz}\right]$ form a base of the left null space of the matrix ${\rm\bf col}\!\left\{ U_{E2}^{T} B_{\rm\bf x},\; B_{\rm\bf z} \right\}$. It is not difficult to see that
\begin{equation}
{\rm\bf span}\left(
\left[\begin{array}{c} N_{\rm\bf Bx}^{T} \\
N_{\rm\bf Bz}^{T}
\end{array}\right]\right) =
{\rm\bf null}\left(\left[\begin{array}{c} U_{E2}^{T} B_{\rm\bf x} \\ B_{\rm\bf z}
\end{array}\right]^{T}\right)
\label{eqn:a31}
\end{equation}
Hence, the condition of Equation (\ref{eqn:a30}) is equivalent to that the following matrix is of FRR,
\begin{displaymath}
\left[\begin{array}{ccc} 0 & U_{E2}^{T}A_{\rm\bf xv} & U_{E2}^{T}B_{\rm\bf x} \\
I & A_{\rm\bf zv} V_{E2} &  B_{\rm\bf z}
\end{array}\right]
\end{displaymath}
The latter is obviously equal to that the matrix $U_{E2}^{T} \left[\begin{array}{cc}
A_{\rm\bf xv} & B_{\rm\bf x} \end{array}\right]$ is of FRR, which is further equivalent to
\begin{equation}
{\rm\bf null}\left(
\left[\begin{array}{cc}
A_{\rm\bf xv} & B_{\rm\bf x} \end{array}\right]^{T}
\right)
\bigcap {\rm\bf span}\left( U_{E2} \right)=\left\{\; 0 \; \right\}
\label{eqn:a32}
\end{equation}
as the matrix $U_{E2}$ is of FCR from its definitions.

Note that $U_{E}=\left[U_{E1} \;\; U_{E2} \right]$ is an orthogonal matrix. Straightforward matrix operations show that ${\rm\bf span}\left(U_{E2}\right) = {\rm\bf null}\left(U_{E1}^{T}\right)$. Hence, the matrix $U_{E2}^{T} \left[\begin{array}{cc}
A_{\rm\bf xv} & B_{\rm\bf x} \end{array}\right]$ is of FRR, if and only if the matrix $\left[\begin{array}{ccc} U_{E1} &
A_{\rm\bf xv} & B_{\rm\bf x} \end{array}\right]$ holds such a characteristics.
As the matrices $A_{\rm\bf xv}$, $B_{\rm\bf x}$ and $U_{E1}$ are consistently block diagonal, it can be claimed directly from Lemma \ref{lemma:3} that the matrix $\left[\begin{array}{ccc} U_{E1} &
A_{\rm\bf xv} & B_{\rm\bf x} \end{array}\right]$ is of FRR, if and only if for each $i\in\{1,2,\cdots,N\}$, the matrix $\left[\begin{array}{ccc} U_{E1}(i) &
A_{\rm\bf xv}(i) & B_{\rm\bf x}(i) \end{array}\right]$ meets this requirement.

The proof can now be completed using Lemma \ref{lemma:4}.   \hspace{\fill}$\Diamond$

\vspace{0.5cm}
\hspace*{-0.45cm}{\rm\bf Proof of Corollary \ref{corollary:1}:}
From the proof of Theorem \ref{theorem:3}, it is clear that there exists a SCM $\Phi$, such that the NDS described by Equations (\ref{eqn:17}) and (\ref{eqn:2}) satisfies Condition I, if and only if the matrix
\begin{equation}
\left[\begin{array}{c}
A_{\rm\bf zx} + B_{\rm\bf z}\left(I_{n_{\rm\bf u}}-FD_{\rm\bf u}\right)^{-1} FC_{\rm\bf x}  \\
U_{E2}^{T}\left[A_{\rm\bf xx} + B_{\rm\bf x}\left(I_{n_{\rm\bf u}}-FD_{\rm\bf u}\right)^{-1} FC_{\rm\bf x} \right]\\
V_{E1}^{T} \end{array}\right]
\label{eqn:a33}
\end{equation}
is of FCR. Note that all the matrices in the above expression, that is, $A_{\rm\bf xx}$, $A_{\rm\bf zx}$, $B_{\rm\bf x}$, $F$ and $U_{E2}$, etc., have a consistent block diagonal structure. It is not difficult to understand that the matrix of Equation (\ref{eqn:a33}) is of FCR, if and only if for each $i=1,2,\cdots,N$, the following matrix holds this property,
\begin{eqnarray}
& &\hspace*{-0.7cm} \left[\!\!\begin{array}{c}
A_{\rm\bf zx}(i) + B_{\rm\bf z}(i)\left(I-F(i)D_{\rm\bf u}(i)\right)^{-1} F(i)C_{\rm\bf x}(i)  \\
U_{E2}^{T}(i)\! \left[A_{\rm\bf xx}(i) \!+\! B_{\rm\bf x}(i)\left(I \!-\! F(i)D_{\rm\bf u}(i)\right)^{-1} F(i)C_{\rm\bf x}(i) \right]\\
V_{E1}^{T}(i) \end{array} \!\!\!\right] \nonumber\\
& &\hspace*{-1.0cm}=\!\!
\left[\!\!\begin{array}{c}
A_{\rm\bf zx}(i)\\
U_{E2}^{T}(i)A_{\rm\bf xx}(i)\\
V_{E1}^{T}(i) \end{array} \!\!\right]
\!+\!
\left[\!\!\begin{array}{c}
B_{\rm\bf z}(i)\\
U_{E2}^{T}(i) B_{\rm\bf x}(i) \\
0 \end{array} \!\!\right]\times \nonumber\\
& & \hspace*{2.0cm}
\left(I-F(i)D_{\rm\bf u}(i)\right)^{-1} F(i)C_{\rm\bf x}(i)
\label{eqn:a34}
\end{eqnarray}
in which the dimension of the identity matrix is equal to $n_{\rm\bf u}(i)$. This dimension is omitted to simplify the expressions.

Assume now that $i\not\in {\cal D}_{I}$. Then according to the definition of the set ${\cal D}_{I}$, it is obvious that $F(i)=0_{n_{\rm\bf u}(i)\times n_{\rm\bf y}(i)}$ makes the left hand side matrix of Equation (\ref{eqn:a34}) have a FCR.

When $i\in {\cal D}_{I}$, using similar arguments as those in the proof of Theorem \ref{theorem:1}, it can be proven that the matrix in the right hand side of Equation (\ref{eqn:a34}) is of FCR, if and only if the following matrix is
\begin{equation}
\left[\!\!\begin{array}{cc}
A_{\rm\bf zx}(i)  &   B_{\rm\bf z}(i) \\
U_{E2}^{T}(i)A_{\rm\bf xx}(i)& U_{E2}^{T}(i) B_{\rm\bf x}(i) \\
V_{E1}^{T}(i) & 0 \\
-F(i)C_{\rm\bf x}(i) & I-F(i)D_{\rm\bf u}(i)
\end{array} \!\!\right]
\label{eqn:a35}
\end{equation}

On the other hand, from the definition of the set ${\cal D}_{I}$, we have that the matrix ${\rm\bf col}\left\{ U_{E2}^{T}(i)A_{\rm\bf xx}(i),\; A_{\rm\bf zx}(i),\;
V_{E1}^{T}(i) \right\}$ is not of FCR. It can therefore be declared that  the following matrix
\begin{equation}
\left[\!\!\begin{array}{cc}
A_{\rm\bf zx}(i)  &   B_{\rm\bf z}(i) \\
U_{E2}^{T}(i)A_{\rm\bf xx}(i)& U_{E2}^{T}(i) B_{\rm\bf x}(i) \\
V_{E1}^{T}(i) & 0
\end{array} \!\!\right]
\label{eqn:a36}
\end{equation}
is not of FCR, also, which means that the right null space of this matrix contains a nonzero vector. That is, the matrix
\begin{displaymath}
\left[\!\!\begin{array}{cc}
A_{\rm\bf zx}(i)  &   B_{\rm\bf z}(i) \\
U_{E2}^{T}(i)A_{\rm\bf xx}(i)& U_{E2}^{T}(i) B_{\rm\bf x}(i) \\
V_{E1}^{T}(i) & 0
\end{array} \!\!\right]^{\perp}
\end{displaymath}
is not equal to a zero matrix. Partition it as
\begin{equation}
\left[\!\!\begin{array}{cc}
A_{\rm\bf zx}(i)  &   B_{\rm\bf z}(i) \\
U_{E2}^{T}(i)A_{\rm\bf xx}(i)& U_{E2}^{T}(i) B_{\rm\bf x}(i) \\
V_{E1}^{T}(i) & 0
\end{array} \!\!\right]^{\perp}
=
\left[\!\!\begin{array}{c}
N_{\rm\bf x}(i)   \\
N_{\rm\bf u}(i)\end{array} \!\! \right]
\label{eqn:a38}
\end{equation}
in which the matrices $N_{\rm\bf x}(i)$ and $N_{\rm\bf u}(i)$ have respectively $n_{\rm\bf x}(i)$ and $n_{\rm\bf u}(i)$ rows. Then, it can be straightforwardly shown that the matrix of Equation (\ref{eqn:a35}) is of FCR, if and only if the matrix $\left[ -F(i)C_{\rm\bf x}(i)\;\;  I-F(i)D_{\rm\bf u}(i)
\right]\left[\!\!\begin{array}{c}
N_{\rm\bf x}(i)   \\
N_{\rm\bf u}(i)\end{array} \!\! \right]$ is of FCR.

Note that
\begin{eqnarray}
& &\!\!\!\! \left[ -F(i)C_{\rm\bf x}(i)\;\;  I-F(i)D_{\rm\bf u}(i)
\right]\left[\!\!\begin{array}{c}
N_{\rm\bf x}(i)   \\
N_{\rm\bf u}(i)\end{array} \!\! \right] \nonumber\\
&=&\!\!\!\! -N_{\rm\bf u}(i) +
F(i)[C_{\rm\bf x}(i)N_{\rm\bf x}(i)+D_{\rm\bf u}(i)N_{\rm\bf u}(i)]
\label{eqn:a37}
\end{eqnarray}
It can therefore be declared from Lemma \ref{lemma:2} that there exists a $F(i)$, such that the matrix of Equation (\ref{eqn:a35}) is of FCR, if and only if the matrix $\left[\!\!\begin{array}{c}
N_{\rm\bf u}(i)   \\
C_{\rm\bf x}(i)N_{\rm\bf x}(i)+D_{\rm\bf u}(i)N_{\rm\bf u}(i)\end{array} \!\! \right]$ is of FCR.

On the other hand, note that the matrix ${\rm\bf col}\{N_{\rm\bf x}(i),\; N_{\rm\bf u}(i)\}$ is of FCR from its definition. Moreover,
\begin{displaymath}
\left[\!\!\begin{array}{c}
N_{\rm\bf u}(i)   \\
C_{\rm\bf x}(i)N_{\rm\bf x}(i)+D_{\rm\bf u}(i)N_{\rm\bf u}(i)\end{array} \!\! \right]
=
\left[\!\!\begin{array}{cc}
C_{\rm\bf x}(i)  &   D_{\rm\bf u}(i) \\
0 & I
\end{array} \!\!\right]\!\!
\left[\!\!\begin{array}{c}
N_{\rm\bf x}(i)   \\
N_{\rm\bf u}(i)\end{array} \!\! \right]
\end{displaymath}
We therefore have that the left hand side matrix in the above equality is of FCR, if and only if
\begin{equation}
{\rm\bf null}\left(\left[\!\!\begin{array}{cc}
C_{\rm\bf x}(i)  &   D_{\rm\bf u}(i) \\
0 & I
\end{array} \!\!\right]\right)\bigcap{\rm\bf span}\left(
\left[\!\!\begin{array}{c}
N_{\rm\bf x}(i)   \\
N_{\rm\bf u}(i)\end{array} \!\! \right]\right)=\{0\}
\label{eqn:a39}
\end{equation}

On the basis of Equation (\ref{eqn:a38}), direct matrix operations show that the condition of Equation (\ref{eqn:a39}) is equivalent to that the matrix
\begin{displaymath}
\left[\!\!\begin{array}{cc}
A_{\rm\bf zx}(i)  &   B_{\rm\bf z}(i) \\
U_{E2}^{T}(i)A_{\rm\bf xx}(i)& U_{E2}^{T}(i) B_{\rm\bf x}(i) \\
V_{E1}^{T}(i) & 0 \\
C_{\rm\bf x}(i)  &   D_{\rm\bf u}(i) \\
0 & I
\end{array} \!\!\right]
\end{displaymath}
is of FCR, which is further equivalent to that the matrix ${\rm\bf col}\left\{ U_{E2}^{T}(i)A_{\rm\bf xx}(i),\; A_{\rm\bf zx}(i),\; C_{\rm\bf x}(i),\;
V_{E1}^{T}(i) \right\}$ is of FCR. This completes the proof.   \hspace{\fill}$\Diamond$

\vspace{0.5cm}
\hspace*{-0.45cm}{\rm\bf Proof of Corollary \ref{corollary:2}:}
Equation (\ref{eqn:17}), together with the proof of Theorem \ref{theorem:4}, makes it is clear that in order to guarantee the existence of a SCM $\Phi$, such that Condition II is satisfied by the NDS of Equations (\ref{eqn:1}), (\ref{eqn:2}) and (\ref{eqn:13}), if and only if the matrix $\left[ U_{E1}\;\; A_{\rm\bf xv} \!+\! B_{\rm\bf x}\left(I_{n_{\rm\bf u}} \!-\! FD_{\rm\bf u}\right)^{-1}\!\! FC_{\rm\bf v}\;\;
B_{\rm\bf x}\left(I_{n_{\rm\bf u}} \!-\! FD_{\rm\bf u}\right)^{-1} \!\!F \right]$ is of FRR, which is equivalent to that the matrix
\begin{equation}
\left[\begin{array}{c}
A_{\rm\bf xv}^{T} + C_{\rm\bf v}^{T}F^{T}\left(I_{n_{\rm\bf u}}-D_{\rm\bf u}^{T}F^{T}\right)^{-1} B_{\rm\bf x}^{T}  \\
F^{T}\left(I_{n_{\rm\bf u}}-D_{\rm\bf u}^{T}F^{T}\right)^{-1} B_{\rm\bf x}^{T}\\
U_{E1}^{T} \end{array}\right]
\label{eqn:a40}
\end{equation}
is of FCR. Once again, note that all the matrices in the above expression, that is, $A_{\rm\bf xv}$, $C_{\rm\bf v}$, $B_{\rm\bf x}$, $D_{\rm\bf u}$, $F$ and $U_{E1}$, have a consistent block diagonal structure. It is easy to see that the matrix in Equation (\ref{eqn:a40}) is of FCR, if and only if for each $i=1,2,\cdots,N$, the following matrix has this property,
\begin{eqnarray}
& &\hspace*{-0.5cm}
\left[\!\!\!\! \begin{array}{c}
A_{\rm\bf xv}^{T}(i) + C_{\rm\bf v}^{T}(i)F^{T}(i)\left(I_{n_{\rm\bf u}(i)}-D_{\rm\bf u}^{T}(i)F^{T}(i)\right)^{-1} B_{\rm\bf x}^{T}(i)  \\
F^{T}(i)\left(I_{n_{\rm\bf u}(i)}-D_{\rm\bf u}^{T}(i)F^{T}(i)\right)^{-1} B_{\rm\bf x}^{T}\\
U_{E1}^{T} \end{array}\!\!\!\!\right]    \nonumber\\
& &\hspace*{-0.8cm}=\!\!
\left[\!\!\begin{array}{c}
A_{\rm\bf xv}^{T}(i)\\
0_{{n_{\rm\bf y}(i)}\times {n_{\rm\bf e}(i)}}\\
U_{E1}^{T}(i) \end{array} \!\!\right]
\!+\!
\left[\!\!\begin{array}{c}
C_{\rm\bf v}^{T}(i)\\
I_{n_{\rm\bf y}(i)} \\
0_{{r(i)}\times {n_{\rm\bf y}(i)}} \end{array} \!\!\right]\times \nonumber\\
& & \hspace*{1.5cm}
\left(I_{n_{\rm\bf y}(i)}-F^{T}(i)D_{\rm\bf u}^{T}(i)\right)^{-1} F^{T}(i)B_{\rm\bf x}^{T}(i)
\label{eqn:a41}
\end{eqnarray}

Note that when the matrix $\left[ U_{E1}(i)\; A_{\rm\bf xv}(i)\; B_{\rm\bf x}(i) \right]$ is not of FRR, it is certain that the matrix $\left[ U_{E1}(i)\; A_{\rm\bf xv}(i) \right]$ is also not of FRR, which further means that the transpose of the latter is not of FCR. This implies that if $i\in{\cal D}_{II}$, then the following matrix is certainly not of FCR,
\begin{displaymath}
\left[\!\!\begin{array}{cc}
A_{\rm\bf xv}^{T}(i)  & C_{\rm\bf v}^{T}(i) \\
0_{{n_{\rm\bf y}(i)}\times {n_{\rm\bf e}(i)}} & I_{n_{\rm\bf y}(i)} \\
U_{E1}^{T}(i) & 0_{{r(i)}\times {n_{\rm\bf y}(i)}}\end{array} \!\!\right]
\end{displaymath}
In addition, note that the right hand side of Equation (\ref{eqn:a41}) has completely the same form as that of Equation (\ref{eqn:a34}). Similar arguments as those in the proof of Corollary \ref{corollary:1} show that there exists a matrix $F(i)$ with $i\in{\cal D}_{II}$, such that the left hand side matrix of Equation (\ref{eqn:a41}) is of FCR, if and only if the matrix ${\rm\bf col}\{ A_{\rm\bf xv}^{T}(i),\; 0_{{n_{\rm\bf y}(i)}\times {n_{\rm\bf e}(i)}},\; U_{E1}^{T}(i),\; B_{\rm\bf x}^{T}(i) \}$ is of FCR, which is equivalent to that the matrix $\left[ U_{E1}(i),\;  A_{\rm\bf xv}(i),\; B_{\rm\bf x}(i) \right]$ is of FRR. The latter is clearly a contradiction to the assumption that $i\in{\cal D}_{II}$, and implies that when the set ${\cal D}_{II}$ is not empty, the  static feedback of Equation (\ref{eqn:13}) is not helpful in constructing an NDS satisfying Condition II.

This completes the proof.   \hspace{\fill}$\Diamond$

\end{document}